\documentclass[aps,10pt,prd,preprintnumbers,amsmath,amssymb,floatfix,nofootinbib]{revtex4}
\usepackage{epsfig}
\usepackage{amsfonts}
\usepackage{amsmath}
\usepackage{amssymb}
\usepackage{slashed}
\usepackage{color}

\newcommand{\ep}{\epsilon}

\newcommand{\al}{\alpha}
\newcommand{\bt}{\beta}
\newcommand{\g}{\gamma}

\newcommand{\Fmu}{F_{\mu\nu}}

\newcommand{\Nb}{\bar N}
\newcommand{\Fp}{F_\pi}

\newcommand{\mpi}{m_{\pi}}

\newcommand{\dslash}[1]{#1 \llap{/\kern-0.5pt}}
\newcommand{\Dslash}[1]{#1 \llap{/\kern+1.2pt}}
\newcommand{\DDslash}[1]{#1 \llap{/\kern+2.3pt}}
\newcommand{\dslashh}[1]{#1 \llap{/\kern+1pt}}

\newcommand{\boldtau}{\mbox{\boldmath $\tau$}}
\newcommand{\boldpi}{\mbox{\boldmath $\pi$}}

\def\gpbz{{\bar g}_\pi^{(0)}}
\def\gpbo{{\bar g}_\pi^{(1)}}

\newcommand{\lsim}{\buildrel < \over {_\sim}}

\newcommand{\be}{\begin{equation}}
\newcommand{\ee}{\end{equation}}
\newcommand{\bee}{\begin{equation*}}
\newcommand{\eee}{\end{equation*}}
\newcommand{\bea}{\begin{eqnarray}}
\newcommand{\eea}{\end{eqnarray}}
\newcommand{\bean}{\begin{eqnarray*}}
\newcommand{\eean}{\end{eqnarray*}}

\begin{document}

\preprint{ACFI-T14-01}

\title{Nucleon Electric Dipole Moments and the Isovector \\Parity- and Time-Reversal-Odd
Pion-Nucleon Coupling}

\author{Chien-Yeah Seng$^{1}$}
\author{Jordy de Vries$^2$}
\author{Emanuele Mereghetti$^{3}$}
\author{Hiren H. Patel$^{4}$}
\author{Michael Ramsey-Musolf$^{1,5}$}\affiliation{\hspace{5mm}
\\
$^{1}${\it Amherst Center for Fundamental Interactions\\
Department of Physics, University of Massachusetts Amherst\\
Amherst, MA 01003 USA}\\
\\$^{2}${\it Institute for Advanced Simulation, Institut f\"ur Kernphysik,
and J\"ulich Center for Hadron Physics, Forschungszentrum J\"ulich,
D-52425 J\"ulich, Germany}\\
\\
$^{3}${\it Ernest Orlando Lawrence Berkeley National Laboratory,
University of California, \\
Berkeley, CA 94720, USA}\\
\\
$^{4}$ {\it Particle and Astro-Particle Physics Division \\
Max-Planck Institute for Nuclear Physics {\rm{(MPIK)}} \\
Saupfercheckweg 1, 69117 Heidelberg, Germany}\\
\\
$^{5}${\it Kellogg Radiation Laboratory, California Institute of Technology\\
Pasadena, CA 91125 USA}\\
}

\date{\today}

\begin{abstract}
The isovector time-reversal- and parity-violating pion-nucleon coupling $\gpbo$ is uniquely sensitive to
dimension six interactions between right-handed light quarks and the Standard Model Higgs doublet that naturally
arises in left-right symmetric models. Recent work has used the $\gpbo$-induced one-loop contribution to the neutron electric dipole moment $d_n$, together with the present experimental $d_n$ bound, to constrain the CP-violating parameters of the left-right symmetric model. We show that this and related analyses are based on an earlier meson theory $d_n$ computation that is not consistent with the power-counting appropriate for an effective field theory. We repeat the one-loop calculation using heavy baryon chiral perturbation theory and find that the resulting $d_n$ sensitivity to $\gpbo$ is suppressed, implying more relaxed constraints on the parameter space of the left-right symmetric model. Assuming no cancellations between this loop contribution and other contributions, such as the leading order EDM low-energy constant,
the present limit on $d_n$ implies $|\gpbo| \lsim 1.1\times 10^{-10}$.
\end{abstract}

\pacs{14.20.Dh,13.75.Gx,11.30.-j,13.40.Em}
\maketitle

\section{Introduction}\label{sec2}
Electric dipole moments (EDMs) of neutral atoms, molecules, and the neutron provide sensitive probes
of new sources of time-reversal (T) and parity (P) violation.
Current upper limits on the EDMs of the $^{199}$Hg atom \cite{Griffith:2009zz}, $d_A(^{199}\mathrm{Hg})$, and neutron \cite{Baker:2006ts}, $d_n$, place tight constraints on
the QCD vacuum angle within the Standard Model (SM) as well as on possible sources of
sources of CP-violation (CPV) arising from physics beyond the SM (BSM)\footnote{In what follows, we assume CPT conservation, so that the signal for a non-vanishing neutron EDM would also indicate the presence of CP-violation.}.  The existence of BSM CPV is needed in order to explain the cosmic baryon asymmetry  (for a recent review, see Ref.~\cite{Morrissey:2012db}).
If the asymmetry had been generated at temperatures of order the electroweak (EW) scale, then $d_n$ provides a particularly sensitive probe.

At energies below the scale of BSM interactions $\Lambda$ but above the EW scale, one may characterize the effects of BSM CPV interactions in terms of an effective theory involving only SM fields:
\be
\label{eq:LCPV1}
\mathcal{L}_\mathrm{CPV} = \mathcal{L}_\mathrm{CKM}+\mathcal{L}_{\bar\theta}
+\mathcal{L}_\mathrm{BSM}^\mathrm{eff}\ ,
\ee
where $\mathcal{L}_\mathrm{CKM}$ and $\mathcal{L}_{\bar\theta}$ denote the SM Cabibbo-Kobayashi-Maskawa (CKM)~\cite{Kobayashi:1973fv} and QCD vacuum angle \cite{'tHooft:1976up, Jackiw, Callan} CPV Lagrangians, respectively, and
\be
\label{eq:LCPV2}
\mathcal{L}_\mathrm{BSM}^\mathrm{eff}= \frac{1}{\Lambda^2}\ \sum_i
\alpha^{(6)}_i \, \mathcal{O}_i^{(6)} +\cdots\ ,
\ee
gives the set of non-renormalizable CPV effective
operators at the weak scale $v=246$ GeV generated by BSM physics at a scale
$\Lambda>v$.  For brevity, we have indicated only those entering at dimension ($d$) six, while the $+\cdots$ indicate those of higher dimension\footnote{
A complete list of dimension six operators may be found in Refs.~\cite{Grzadkowski:2010es}, while those directly relevant to EDMs are listed in Tables
three and four of Ref.~\cite{Engel:2013lsa}.}. Among the more widely considered $d=6$ CPV operators are the elementary fermion EDMs, the quark chromo-EDMs, and the Weinberg three-gluon operator.

In this study, we focus on one particular $d=6$ operator that naturally arises in left-right symmetric model (LRSM) extensions of the SM, that gives rise to EDMs of nucleons, nuclei, and diamagnetic atoms, and that has received considerably less attention than the other operators that arise at this order in the effective theory:
\be
\mathcal{O}_{\varphi ud} = i ({\tilde\varphi}^\dag D_\mu \varphi) {\bar u}_R \gamma^\mu d_R\ ,
\ee
where $\varphi$ is the Higgs doublet, ${\tilde\varphi}=i\tau_2\varphi^\ast$, $D_\mu$ is the SU(2)$_L\times$U(1)$_Y$ covariant derivative and $u_R$ ($d_R$) is the right-handed up-quark (down-quark) field. In LRSMs, the corresponding Wilson coefficient $C_{\varphi u d}$ is generated by mixing between the left- and right-handed $W$ bosons in the presence of either spontaenous CPV and/or explicit CPV in the right-handed quark CKM matrix.

After electroweak symmetry-breaking wherein $\varphi^T\to (0, v/\sqrt{2})$, the exchange of the $W^\pm$ contained in the covariant derivative with a left-handed quark current leads to an effective four quark interaction\footnote{Corrections due to the evolution of the four quark interaction to hadronic scales are minor, see the discussion in Ref. \cite{Dekens:2013zca}. }
\be
\label{eq:lreff0}
\mathcal{L}^\mathrm{eff}_\mathrm{LR,\, CPV} =
-i \frac{\mathrm{Im}\, C_{\varphi u d}}{\Lambda^2}
\left[{\bar d}_L\gamma^\mu u_L\, {\bar u}_R\gamma_\mu d_R
-{\bar u}_L\gamma^\mu d_L\, {\bar d}_R\gamma_\mu u_R\right] \ .
\ee
The interaction in Eq.~(\ref {eq:lreff0}) breaks isospin symmetry,
thereby giving rise to, among other interactions, the isovector TVPV $\pi NN$ interaction:
\be
\label{eq:piNN1}
\mathcal{L}_{\pi N,\ I=1}^\mathrm{TVPV} = \gpbo {\bar N}\pi^0 N\ ,
\ee
where $N$ and $\pi^0$ are the nucleon and neutral pion fields, respectively. This interaction leads to long-range contributions to the  nuclear Schiff moment that induces $d_A(^{199}\mathrm{Hg})$ as well as long-range contributions to $d_n$ that can be computed in chiral perturbation theory. The present limits on these EDMs, thus, imply constraints on the mass $M_{W_R}$ of the right-handed $W$-boson and associated CPV phases in the LRSM.

Following this line of reasoning, the authors of Refs.~\cite{Zhang:2007da, ji}
have derived constraints on  $M_{W_R}$ and the strength of spontaneous CPV in the LRSM  from the limits on $d_n$ and the corresponding implications of CPV in the neutral kaon sector. The results imply that $M_{W_R}> 10$ TeV. In related work, the authors of Ref.~\cite{Ng:2011ui} observed that $\mathcal{O}_{\varphi ud}$ will also induce a semi-leptonic CPV operator that contributes to neutron decay. Even without specifying to the LRSM, the $d_n$ limits on $C_{\varphi u d}$ then constrain the magnitude of possible effects in T-odd neutron decay correlations. In both cases, the $d_n$ constraints relied on an earlier pion-loop calculation performed by the authors of Ref.~\cite{He:1992db} using a relativistic meson-nucleon field theory approach. The results indicate that the leading term in $d_n$ resulting from the interaction (\ref{eq:piNN1}) is proportional to the neutron anomalous magnetic moment $\kappa_n$ and is independent of the pion-to-nucleon mass ratio, $m_\pi/m_N$. From the standpoint of effective field theory (EFT), this result is surprising, as the anomalous magnetic moment vertex brings in an inverse power of the nucleon mass while consistent power counting in chiral perturbation theory requires that loops only bring in momenta of order of the pion mass. The absence of any $m_\pi/m_N$ suppression in the computation of Ref.~\cite{He:1992db} is not consistent with this expectation.

In what follows, we repeat the pion loop computation associated with (\ref{eq:piNN1}) using heavy baryon chiral perturbation theory (HBChPT)~\cite{Jenkins:1990jv} and show that the result proportional to $\gpbo\kappa_n$ is suppressed by $(m_\pi/m_N)^2\sim 0.02$. HBChPT implements the power counting required by an EFT by expanding about both the static nucleon ($m_N\to\infty$) and chiral ($m_\pi\to 0$) limits. Our results imply considerably weaker constraints on $C_{\varphi u d}$ from the long-range contribution to $d_n$ than obtained in the studies of Refs.~\cite{Zhang:2007da,ji,Ng:2011ui}. Presently uncalculable short-distance contributions associated with loop momenta of order one GeV that reside in the nucleon EDM counterterm may imply stronger constraints as suggested by na\"ive dimensional analysis (NDA).  In this context, one may view the relativistic meson theory computation of Ref.~\cite{He:1992db}  as providing a model estimate of the short-distance contributions. Generally speaking, however, both the sign and magnitude of NDA and/or model estimates for the short distance contributions are subject to uncertainty, so the most conservative  implications will be inferred from the calculable long-distance terms.

In this respect, we note that the diamagnetic EDMs provide an in principle more robust benchmark than $d_n$, as the nuclear Schiff moment arises from tree-level pion exchange between two nucleons and is relatively free from the uncertainties associated with short-distance contributions. In practice, the many-body nuclear theory uncertainty associated with the interaction (\ref{eq:piNN1}) are substantial \cite{Engel:2013lsa}, with the situation for $^{199}$Hg being particularly unsettled. Looking to the future, storage-ring searches for  EDMs of light nuclei such as the deuteron or ${}^3$He nucleus \cite{Pretz:2013us} would provide theoretically cleaner probes of $\mathcal{O}_{\varphi ud}$ since the short-distance contributions to such EDMs are relatively minor and since the few-body nuclear theory is well under control \cite{deVries:2011an}.  In the immediate term, however,  the long-range contribution to $d_n$ appears to be the most trustworthy avenue for accessing $\mathcal{O}_{\varphi ud}$.

In the remainder of this paper, we discuss the details of our calculation. In Section \ref{sec:hbchpt} we summarize the HBChPT framework as it applies to the computation of TVPV observables and give the details of our nucleon EDM computation. In Section \ref{sec:compare} we compare our results with those of Ref.~\cite{He:1992db}. We discuss the implications and summarize in Section \ref{sec:imply}.

\section{HBChPT Calculation}\label{sec:hbchpt}

Loop computations performed with a relativistic meson-nucleon field theory and dimensional regularization include explicit contributions from loop momenta of order $m_N$, thereby eliminating the possibility of a consistent power counting\footnote {A relativistic approach can provide a reliable power counting if more complicated regularizations schemes are applied, for a review see Ref. \cite{Bernard}.}.
 In HBChPT \cite{Jenkins:1990jv}, one removes these contributions by first redefining the nucleon degrees of freedom in terms of heavy fields having fixed velocity $v$
\begin{eqnarray}\label{Nv}
N_v = \frac{1+\dslash v}{2}e^{i m_N v\cdot x} N\ ,
\end{eqnarray}
where
\begin{eqnarray}\label{decomposition}
p^\mu = m_N v^\mu + k^\mu\ ,
\end{eqnarray}
with $k$ being a residual momentum. We henceforth omit the \lq\lq $v$" subscript. Derivatives acting on the heavy fields give the small residual momenta, and the propagator of a heavy-nucleon field no longer contains the nucleon mass.
The results of loop integrals involving the $N$ fields then scale with powers of $Q/m_N$ and $Q/\Lambda_\chi$, where $Q$ is of order $m_\pi$ or the external momentum (assumed to be well below one GeV), $\Lambda_\chi=2\pi F_\pi$ is the scale of chiral symmetry breaking, and $F_\pi=186$ MeV is the pion decay constant\footnote{Note that other work in HBChPT uses $f_\pi=F_\pi/2$.}. One, thus, obtains a consistent EFT power counting.

The HBChPT interactions are constructed from the heavy nucleon and
pion fields, the velocity $v^\mu$,  and the spin $S^\mu$ with
$S=(\vec{\sigma}/2, 0)$ in the nucleon rest frame $v=(\vec{0}, 1)$.
It is also useful to project vectors in their components parallel
and orthogonal to the velocity. We use a subscript $\perp$ to denote
the perpendicular components. For example, the perpendicular
covariant derivative is
\begin{equation}
\mathcal D^{\mu}_{\perp} =\mathcal D^{\mu} - v^{\mu} v \cdot \mathcal D\ .
\end{equation}
We note that HBChPT operators are manifestly invariant under
rotations and translations.
Under boosts, the HBChPT operators are only invariant order by order in the $Q/m_N$ expansion, leading to non-trivial constraints on the interactions in the effective Lagrangian and on their coefficients
\cite{Jenkins:1990jv, ManoharLuke}.

The resulting, leading order (LO)  $P$- and $T$-conserving  Lagrangian for $f$ fermion fields is (see Ref. \cite{reviewchiPT} for more details)
\begin{equation}
{\mathcal L}^{(0)} =
\frac{1}{2} D_{\mu} \boldpi \cdot D^{\mu} \boldpi - \frac{\mpi^2}{2D}\boldpi^2
+ \bar{N}\left( iv\cdot {\mathcal D}
         -\frac{2 g_A}{F_\pi}S^{\mu} \boldtau\cdot D_{\mu} \boldpi\right) N\ ,
\label{LagrCons}
\end{equation}
where $g_A = 1.27$ is the pion-nucleon axial-vector coupling.
At LO, electromagnetism can be included by making the derivatives in Eq. \eqref{LagrCons} covariant under $U_{\textrm{em}}(1)$,
\begin{eqnarray}\label{eq:minimal.1}
  (D_{\mu} \pi)_i & \rightarrow & (D_{\mu,\mathrm{em}} \pi)_i =
\frac{1}{D} \left(\partial_{\mu}\delta_{ij} + eA_{\mu}\ep_{3ij}\right)
\pi_j \ ,
\nonumber \\
 \mathcal D_{\mu} N   &\rightarrow & \mathcal D_{\mu,\mathrm{em}} N =
\left[\partial_{\mu}+\frac{i}{F_\pi^2}\boldtau \cdot
\left(\boldpi\times D_{\mu,\mathrm{em}}\boldpi \right)
+ \frac{ie}{2} A_{\mu} \left(1 + \tau_3 \right)\right]N\ ,
\end{eqnarray}
where $e$ is the proton charge, $e>0$ and $D= 1+\boldpi^2/\Fp^2$.
For brevity, in the following we omit the label ``em''.

The anomalous magnetic moment that is of particular interest to this work first appears at next-to-leading order (NLO) in $Q/m_N$, together with the nucleon kinetic energy, a recoil correction to the axial pion-nucleon coupling, the nucleon sigma term, and the proton-neutron mass difference
\begin{eqnarray}
\mathcal{L}^{(1)}
&=& -\frac{1}{2m_N} \bar{N} \mathcal D_{\perp}^{2}N
+\frac{ g_{A}}{F_{\pi} m_N} \left(i v\cdot D \boldpi\right) \cdot
\bar{N}\boldtau\, S\cdot \mathcal D_{-} N  \nonumber
\\ & &+\frac{1}{4 m_N}\varepsilon^{\al\bt\mu\nu}v_\al \Nb
S_\bt\left[(1+\kappa_0)+(1+\kappa_1)\tau_3 \right] N e\Fmu\nonumber
\\ & &
+ \Delta m_N \left(1-\frac{2\boldpi^2}{\Fp^2 D}\right) \Nb N + \frac{\delta m_N}{2} \Nb \left(\tau^3 -\frac{2\pi_3}{\Fp^2 D}\boldtau\cdot\boldpi\right)N+ \ldots\ .
\label{LagStrong1}
\end{eqnarray}
Here, $\kappa_1$ and $\kappa_0$ are  the isovector and isoscalar anomalous magnetic moments (AMMs),   $\kappa_1 = 3.7$, $\kappa_0 = -0.12$, and $\Delta m_N$ and $\delta m_N$ the nucleon sigma term and nucleon mass splitting originating in the quark masses. A lattice calculation found $\delta m_N = 2.26$ MeV~\cite{latticedeltamN}, while evaluations of $\Delta m_N$ range between 45 and 60 MeV~\cite{Gasser:1990ce}. The \lq\lq $+\cdots$" in Eq. \eqref{LagStrong1} denote other operators with chiral index $\Delta =1$, like the pion mass splitting term, which are not relevant for our discussion.

In Eq. \eqref{LagStrong1}, the nucleon AMM couples to the magnetic field, and this interaction does not contribute directly to the nucleon EDM.
However, the AMM does couple to the electric field at next-to-next-to-leading order (NNLO) through the spin-orbit interaction that appears in  $\Delta = 2$ Lagrangian:
\begin{eqnarray}\label{em2}
\mathcal{L}^{(2)}&=&
 \frac{g_A}{4 m_N^2 F_{\pi} } D_{\mu} \boldpi  \cdot  \left(
\bar N \boldtau S^{\mu} \mathcal D^2_{\perp,\, -}  N
-
\bar N \boldtau \mathcal D^{\mu}_{\perp,\, -}
S \cdot \mathcal D_{\perp,\, -}N  \right) \nonumber\\
& &
  -\frac{i}{16 m_N^2} \varepsilon^{\al\bt\mu\nu}\Nb S_\al \left[(1+2\kappa_0)+(1+2\kappa_1)\tau_3\right]\mathcal D_{\bt,\perp,-}
  ]N\,e\Fmu + \ldots\ .
\end{eqnarray}
The first term in Eq. \eqref{em2} is a relativistic correction to
$g_A$, while the second term is the spin-orbit nucleon-photon
coupling. Reparametrization invariance fixes the
coupling constant of the interactions in Eq. \eqref{em2} in terms of
the couplings in Eqs. \eqref{LagrCons} and \eqref{LagStrong1}
\cite{Jenkins:1990jv, ManoharLuke}.

Next we list the TVPV interactions originating in the four-quark operator in Eq.~\eqref{eq:lreff0}. The construction of these interactions has been discussed in detail in Ref. \cite{Lagdim6}, and here we only summarize the results.  As already mentioned, the isovector TVPV pion-nucleon interaction in Eq.~\eqref{eq:piNN1} with coupling $\gpbo$ is induced at leading order. Additionally, the quantum numbers of the neutral pion allow the existence of a pion tadpole term. This tadpole can be removed from the Lagrangian via a field redefinition of the pion and nucleon fields \cite{BiraEmanuele}.
These field redefinitions induce an additional
 LO contribution to the isovector TVPV pion-nucleon that can be absorbed into
 $\gpbo$. The tadpole removal also induces a contribution to the isoscalar TVPV pion-nucleon interaction
\be
\label{eq:piNN0}
\mathcal{L}_{\pi N,\ I=0}^\mathrm{TVPV} = \gpbz  {\bar N}\boldtau\cdot\boldpi N\ .
\ee
Though formally LO, this contribution
vanishes in the isospin limit $m_u = m_d$ and is suppressed by the ratio
$\gpbz/\gpbo = \delta m_N/(2\Delta m_N)\simeq 0.02$.

We find that there exist no NLO corrections proportional to $\gpbo$ and $\gpbz$, because they depend on structures like $(1\pm
v\!\!\!/)(1\mp v\!\!\!/)$ that vanish. The first non-vanishing
corrections proportional to $\gpbo$ appear at NNLO  and are given by
\begin{eqnarray}
\mathcal L^{\mathrm{TVPV}(\mathrm{NNLO})}_{\pi N,\, I=1}&=&
 -\frac{\gpbo}{4 m_N^2}(D_\nu \pi_3)\Nb[S^\mu,S^\nu]\mathcal D_{\mu,-}N - \frac{\gpbo}{8 m_N^2} { \pi_3}\Nb \mathcal D_{\perp,-}^2 N  \ .
\label{relcorrg1}
\end{eqnarray}
For brevity, we do not list the remaining NLO and NNLO TVPV interactions \cite{Lagdim6} that are not relevant for the present discussion.

Finally, the combination of Eq.~\eqref{eq:lreff0} and electromagnetic interactions gives rise to the EDM interaction
\begin{eqnarray}\label{short}
\mathcal L^\mathrm{TVPV} _{\mathrm{em}} &=&
-2 \Nb (\bar d_0 +\bar d_1\tau^3 )
S^\mu  N\, v^\nu \Fmu+ \ldots\ ,
\end{eqnarray}
where $\bar d_0$ and $\bar d_1$ are, respectively, the isoscalar and isovector short-range contributions to the nucleon EDM and the \lq\lq $+\cdots$" stand for terms with one or more pions or relativistic corrections not relevant for the discussion here.

We now compute the contributions to the nucleon EDM generated by the interactions in Eqs.~(\ref{eq:piNN1},\ref{LagrCons},\ref{LagStrong1}-\ref{short}). In the HBChPT framework, the EDM enters as a TVPV contribution to the nucleon EM current:
\begin{eqnarray}
\label{EDFFdef}
J^\mu_{\mathrm{TVPV}}(q)&=&2i\left(d_0
+ d_1\tau_3\right)
\left(S\cdot q v^\mu-S^\mu v\cdot q
+\ldots
\right)\ ,
\label{current}
\end{eqnarray}
where $q^\mu$ is the four-momentum of the outgoing photon and
\be
\mathcal{L}_{\gamma N}^\mathrm{TVPV}=-J^\mu_\mathrm{TVPV}
A_\mu\ .
\ee
The dots in Eq. \eqref{EDFFdef} denote relativistic corrections to the leading order current. The neutron (proton) EDM is given by $d_0-d_1$ ($d_0+d_1$). For virtual photons, the nucleon EDMs become the electric dipole form factors (see, {\em e.g.}, Refs. \cite{BiraHockings, Vri11a, Mer11}). For purposes of our calculation, however, we focus on the static limits.

Contributions to the nucleon EDM arise from tree-level diagrams involving the short-range interactions in Eq.~\eqref{short} and from one-loop diagrams involving the TVPV pion-nucleon vertices. The largest one-loop diagrams involve the $P$- and $T$-conserving vertices from Eq.~\eqref{LagrCons} only, while corrections appear due to insertions of vertices from Eq.~\eqref{LagStrong1} or \eqref{em2}.

\begin{figure}[t]
\centering
\includegraphics[scale = 0.8]{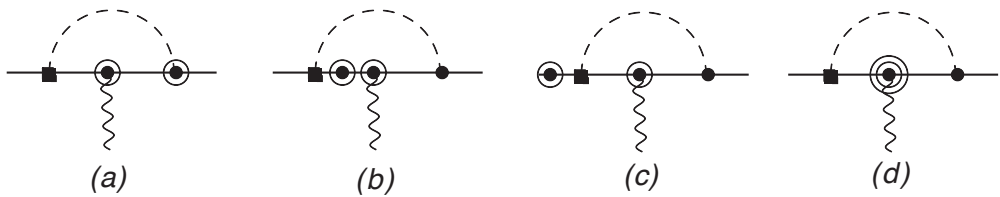}
\caption{One-loop diagrams contributing to the nucleon EDMs at next-to-next-to-leading order. Solid,
dashed and wavy lines represent the propagation of nucleons, pions
and photons, respectively. A square marks the isovector TVPV interaction
from Eq. \eqref{eq:piNN1}, other vertices representing the T- and P-conserving interactions.
Each circle on the vertex represents a suppression
factor proportional to  $1/m_N$. For
simplicity only one possible ordering is shown.}
\label{fig2}
\end{figure}

\begin{figure}[t]
\centering
\includegraphics[scale = 0.8]{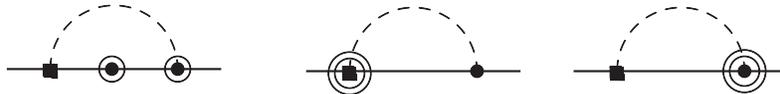}
\caption{One-loop diagrams contributing to wave function renormalization. The notation is as in Fig. \ref{fig2}.}
\label{wave_function}
\end{figure}

The contributions to $d_{0,1}$ proportional to $\gpbo$ and $\gpbz$ up to order $\mathcal{O}(Q/m_N)$ have been computed previously in Refs.~\cite{Mer11, CDVW79}, and they are independent of $\kappa_{0,1}$. The reason being that $\mathcal{O}(Q/m_N)$ electromagnetic interactions given in Eq.~(\ref{LagStrong1})  do not contribute to the nucleon EDM since the interaction with the external field is purely magnetic in the rest frame of the nucleon. The contributions proportional to the AMMs therefore enter at $\mathcal{O}(Q/m_N)^2$ and are generated by the vertex correction diagrams in Fig.~\ref{fig2} along with the corresponding wave function renormalization graphs in Fig. \ref{wave_function}. Because $\gpbz/\gpbo$ is small,
at this order we focus on diagrams involving $\gpbo$ only. The number of open circles indicates the order in the $(Q/m_N)$ expansion. Again, because the $\mathcal O(Q/m_N)$ electromagnetic interactions are purely magnetic, the contributions from the topologies of Figs.~\ref{fig2}(a-c) vanish individually. We also find that the sum of contributions proportional to $\gpbo$ that arise from the $\mathcal{O}(Q/m_N)^2$ wave function renormalization vanish.
The only surviving, non-vanishing contribution is generated by the topology of Fig.~\ref{fig2}(d).

\begin{figure}[t]
\centering
\includegraphics[scale = 0.8]{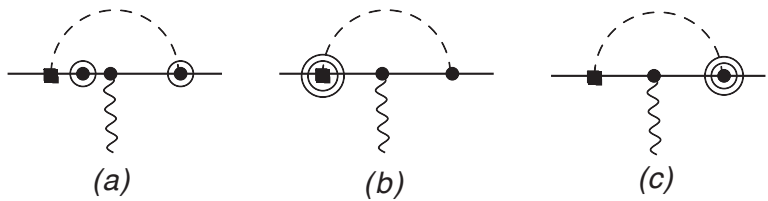}
\caption{Representative one-loop diagrams contributing  at order $(m_\pi/m_N)^2$ to the proton EDM proportional to $\gpbo$ and independent of the anomalous magnetic moment coupling. The notation is as in Fig. \ref{fig2}. }
\label{fig3}
\end{figure}

At $\mathcal{O}(Q/m_N)^2$, an additional set of contributions to the proton EDM proportional to $\gpbo$ are generated by the graphs in Fig.~\ref{fig3}. The latter contain only the coupling to the nucleon charge and various combinations of (a) $\mathcal{O}(Q/m_N)$ corrections to the $\pi NN$ interactions and nucleon propagator as well as  (b) $\mathcal{O}(Q/m_N)^2$ corrections to one of the couplings or propagator and LO interactions for the remaining components of the diagram.
Finally, at this order in principle contributions appear due to two-loop diagrams involving $\gpbo$ and LO $P$- and $T$-conserving vertices. However, the sum of these diagrams vanishes.

Before giving our result, we first quote
 the LO and NLO results, including the contributions involving both $\gpbo$ and $\gpbz$. It should be noted that the results involving $\gpbz$ are actually
 suppressed
by  the smallness of $\gpbz/\gpbo$ for the CPV source in Eq.~(\ref {eq:lreff0}).
The LO contribution proportional to $\gpbz$ has been first calculated in Ref.  \cite{CDVW79}, while the NLO corrections have been calculated in Refs. \cite{Mer11}. Together they give
\begin{eqnarray}\label{result}
d_0^\mathrm{LO+NLO} &=& \bar d_0 - \frac{eg_A}{4\pi^2 F_{\pi}}\;
        \frac{ \pi m_{\pi}}{4m_N}
              \left( 3 \gpbz + \gpbo\right)\ ,
     \nonumber\\
  d_1^\mathrm{LO+NLO} &=& \bar d_1 - \frac{eg_{A}}{4\pi^2 F_{\pi}}
        \left[ \gpbz \left(  L-\ln\frac{m_{\pi}^2}{\mu^2} +\frac{5\pi}{4}\frac{m_\pi}{m_N}  \right)
        +\frac{\pi}{4}\frac{m_\pi}{m_N} \gpbo \right]\ .
\label{eq:lonlo}
\end{eqnarray}
The contribution of $\gpbz$ to the isovector EDM is UV divergent, with divergence encoded in the factor
\begin{eqnarray}
L&\equiv& \frac{2}{4-d}-\g_E+\ln 4\pi \ ,
\end{eqnarray}
in terms of number of space-time dimensions $d$ and the Euler-Mascheroni constant $\g_E$.
The divergence is absorbed by the low-energy constant $\bar d_1$, which also contains an {\em a priori} unknown non-vanishing finite contribution. It is conventional
to absorb $L$ entirely into ${\bar d}_1$, thereby isolating the terms non-analytic in quark mass that are unique to the loops.
Notice that, at this order, the $\gpbo$ terms do not contribute to the neutron EDM. For the interaction $C_{\varphi u d}$ in Eq. \eqref{eq:lreff0}, this implies that the largest non-analytic contributions only affect the proton EDM because $\gpbz$ is suppressed.

At NNLO, $\gpbo$ contributes to the nucleon EDM via
the diagrams in Figs. \ref{fig2} and \ref{fig3}. At this order in
the $Q/m_N$ expansion,
we neglect small $\gpbz m_{\pi}^2/m_N^2$ corrections, and only give terms proportional to $\gpbo$. We find that the contributions are divergent:
\begin{eqnarray}
d_0^{\mathrm{NNLO}} &=& \frac{e g_A\gpbo}{64\pi^2 \Fp}\frac{\mpi^2}{m_N^2}\left[(5+4\kappa_1)\left(L - \log \frac{\mpi^2}{\mu^2} \right)+1\right]\ ,\nonumber\\
d_1^{\mathrm{NNLO}} &=& \frac{e g_A\gpbo}{64\pi^2 \Fp}\frac{\mpi^2}{m_N^2}\left[(5+4\kappa_0)\left(L - \log
\frac{\mpi^2}{\mu^2} \right)+1\right]\ . \label{MDMresult}
\end{eqnarray}
Again, we  can absorb $L$ and other finite pieces analytic
in the quark mass  into the LECs ${\bar d}_{0,1}$. It is also
instructive to express results through $\mathcal{O}(Q/m_N)^2$ for
the neutron and proton EDM's separately: \bea \label{eq:dnfull} d_n
& = & {\bar d}_n-\frac{e g_A}{4\pi^2 F_\pi}\left\{ \gpbz \left( \ln
\frac{m_\pi^2}{m_N^2} -\frac{\pi m_\pi}{2 m_N} \right)
+\frac{\gpbo}{4}\, (\kappa_1-\kappa_0)\frac{m_\pi^2}{m_N^2}\ln  \frac{m_\pi^2}{m_N^2}\right\}\ ,\\
\label{eq:dpfull}
d_p & = & {\bar d}_p+\frac{e g_A}{4\pi^2 F_\pi} \left\{ \gpbz \left( \ln \frac{m_\pi^2}{m_N^2} -\frac{2\pi m_\pi}{m_N} \right) -
\frac{\gpbo}{4}\left[  \frac{2\pi m_\pi}{m_N}  +\left(\frac{5}{2}+\kappa_0+\kappa_1\right) \frac{m_\pi^2}{m_N^2}\ln  \frac{m_\pi^2}{m_N^2}\right]\right\}\ ,
\eea
where we have also absorbed analytic terms into ${\bar d}_p = \bar d_0 + \bar d_1$ (${\bar d}_n = \bar
d_0 - \bar d_1$) and evaluated the result at a renormalization scale
$\mu=m_N$. We observe again that unlike $d_p$, $d_n$ contains no
terms proportional to $\gpbo$ linear in $m_\pi$ and that its
non-analytic quark mass dependence first appears at NNLO. Thus,
compared to the estimates based on Ref.~\cite{He:1992db}, the $d_n$
contributions proportional to $\kappa_{0,1}$ are suppressed by two
powers of $m_\pi/m_N$, leading to a factor of $\sim 50$ suppression.

In principle, the magnitude of the finite parts of the ${\bar d}_{0,1}$, corresponding to \lq\lq short range" dynamics, may be larger than those arising from the loops.
In practice, the relative importance of the short-range  and loop contributions depends on the properties of the four-quark operator in consideration. To illustrate, we consider
the interaction of Eq.~(\ref{eq:lreff0}).  Using general chiral symmetry arguments and NDA \cite{NDA}, we expect~\cite{Lagdim6, Engel:2013lsa}
\bea
\label{eq:nda1}
\gpbo& \sim & \frac{\Lambda_\chi^3}{F_\pi \Lambda^2}\, \frac{\mathrm{Im} C_{\varphi ud}}{(4\pi)^2}\ ,\\
\label{eq:nda2}
{\bar d}_{0,1} & \sim & \frac{e\Lambda_\chi}{\Lambda^2}\, \frac{\mathrm{Im} C_{\varphi ud}}{(4\pi)^2}\ ,
\eea
where $\Lambda_\chi = 2\pi F_\pi\sim m_N$. Substituting the estimate (\ref{eq:nda1}) into Eqs.~(\ref{eq:lonlo}) we have that
\be
\label{NDAestimate}
{\bar d}_{0,1}^{\, \mathrm{NLO,\, loop}} \sim \frac{e}{(2\pi F_\pi)^2}\, \frac{\pi m_\pi}{4m_N}\, \frac{\Lambda_\chi^3}{\Lambda^2}\, \frac{\mathrm{Im} C_{\varphi ud}}{(4\pi)^2} \sim \frac{e\Lambda_\chi}{\Lambda^2}\, \frac{\pi m_\pi}{4m_N}\, \frac{\mathrm{Im} C_{\varphi ud}}{(4\pi)^2}\ .
\ee
The result in Eq.~(\ref{NDAestimate}) is suppressed by one power of $m_\pi/m_N$ with respect to  the NDA estimate (\ref{eq:nda2}).

The non-analytic NNLO loop contributions are suppressed by an additional factor of $ m_{\pi}/m_N$ with respect to  NLO contributions to ${ d}_{0,1}$,
and of $m_{\pi}^2/m_N^2$ with respect to the short-distance low-energy constants $\bar d_{0,1}$.
For the isoscalar EDM, the suppression is somewhat mitigated by the enhancement due to the large value of $\kappa_1$.

To obtain a sense of the relative importance of various terms in Eqs. \eqref{eq:dnfull} and \eqref{eq:dpfull}, we assume $\gpbz/\gpbo = 0.02$ as one expects for the source in Eq.~({\ref{eq:lreff0}). In this case,  the formally LO contribution to $d_n$, proportional to $\gpbz$, and the $\mathcal O(m_\pi^2/m_N^2)$ correction proportional to $\gpbo\times\kappa_n$ are numerically comparable.
For the proton EDM, the NLO contribution proportional to $\gpbo m_{\pi}/m_N$ is about twice as large as the NNLO term. The two contribution enter with opposite sign, and partially cancel, so that net contributions of the two couplings $\gpbz$ and $\gpbo$ to $d_p$ are also of similar magnitude. The comparison with the short-range contributions is more uncertain. If we use the NDA estimates of $\bar d_n$, $\bar d_p$, and $\gpbo$ in Eqs. \eqref{eq:nda1} and \eqref{eq:nda2} we find that the pion loops enter at the $10\%$ to $20\%$ level with respect to the short-range terms. However, deviations from the NDA expectations can affect this conclusion. More robust estimates require a first principles  determination of  $\bar d_{n}$, $\bar d_p$, $\gpbz$ and $\gpbo$ in terms of the coefficient $C_{\varphi u d}$, which, at the moment, is not available (model estimates are summarized in Ref.~\cite{Engel:2013lsa}).

Finally our result for $d_n$ in Eq. \eqref{eq:dnfull} can be used to obtain a limit on $\gpbo$. Using the experimental upper bound $|d_n| < 2.9 \cdot 10^{-26}\,e\,\mathrm{cm}$ \cite{Baker:2006ts} and assuming no cancellations with other contributions, we obtain
\begin{equation}
|\gpbo| \lsim  1.1\times 10^{-10}\ .
\end{equation}
This bound can be used to place an upper bound on the size of the EDMs of the deuteron and Hg, which are particularly sensitive to isospin breaking couplings.
For example, using the results of Ref. \cite{deVries:2011an}, Eq. (26) would limit the deuteron EDM to
$d_d < 10^{-24}$ $ e$~cm.
The bound on $\bar g^{(1)}_\pi$ could be converted into a lower bound on the coefficient of the four-quark operator in Eq.~\eqref{eq:lreff0}. However, since the loop contributes to the nucleon EDM at the 10/20 \% level,  a much more stringent bound on this coefficient can be obtained by the LEC $\bar{d}_0$, and $\bar d_1$, as discussed in the following section.

\section{Comparison with Earlier Work}
\label{sec:compare}

\begin{figure}[t]
\centering
\includegraphics[scale = 0.8]{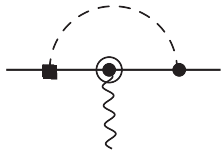}
\caption{One-loop diagram contributing to the nucleon EDM
in relativistic chiral perturbation theory. A square
marks a TVPV interaction, other vertices representing the T- and
P-conserving interactions in Eqs. \eqref{pseudoscalar} and
\eqref{gamma}.} \label{relativistic}
\end{figure}

It is instructive to compare the HBChPT results with those of
Ref.~\cite{He:1992db}. The latter compute the
contribution from Fig.~\ref{relativistic} using the
TVPV interactions  in Eqs.~(\ref{eq:piNN1}) and (\ref{eq:piNN0}),
together with the T- and P-conserving pseudoscalar pion-nucleon coupling
\begin{equation}\label{pseudoscalar}
\mathcal L_p = \frac{2 m_N g_A}{F_{\pi}}\bar N i \gamma^5 \boldtau \cdot \boldpi N\ ,
\end{equation}
the relativistic version of the nucleon propagator, and the nucleon magnetic moment Lagrangian
\begin{equation}\label{gamma}
\mathcal L = -\frac{e}{8m_N} \bar N \sigma^{\mu \nu}
F_{\mu \nu} \left[ \kappa_0 + \kappa_1 \tau_3\right]N \ .
\end{equation}
We have repeated the calculation using these ingredients and obtain
\begin{eqnarray}
d_0 &=&   e(3\gpbz \kappa_0+\gpbo  \kappa_1 ) \,
\frac{g_A}{16\pi^2 F_\pi} F_P\left(\frac{m^2_{\pi}}{m_N^2}\right)\ ,
\nonumber \\
d_1 &=&  e(-\gpbz \kappa_1+\gpbo
\kappa_0 ) \, \frac{g_A}{16\pi^2 F_\pi}
F_P\left(\frac{m^2_{\pi}}{m_N^2}\right)\ ,\label{psEDMres}
\end{eqnarray}
where, when expanded for small $x$,
\begin{equation}\label{FP}
F_P(x) = \frac{3}{2} \left[ \frac{4}{3} - \frac{2}{3}\pi \sqrt{x}  - x \log x + \mathcal O(x^{3/2})\right]\ .
\end{equation}
The result obtained in Ref. \cite{He:1992db} is  similar apart from the
first term in Eq.~\eqref{FP} that is $1$ in their work instead of $4/3$ as we find.
The discrepancy can be traced back to a term containing the product of Dirac matrices $\gamma_\alpha \sigma^{\mu\nu}\gamma^\alpha$. This combination is proportional to $d-4$,  but comes in front of a divergent integral. In the limit $d\rightarrow 4$, the $1/(d-4)$ from the loop integral cancels the $d-4$ in the product of Dirac matrices, leaving a constant piece.


Comparing with the HBChPT results in Eq.~(\ref{MDMresult}), we see
that the relativistic calculation with pseudoscalar coupling yields contributions proportional
to $\gpbo\kappa_{0,1}$ that arise at lower order in $m_\pi/m_N$. The
zeroth order term in Eq.~(\ref{psEDMres}) clearly conflicts with the
power counting expected in a well-behaved EFT  since there appears
no $m_\pi/m_N$ suppression. Using Eq.~(\ref{eq:nda1}) in
Eq.~(\ref{psEDMres}) would conflict with the chiral/NDA expectations
in Eq.~(\ref{NDAestimate}), overestimating the latter by a factor
of $m_N/m_\pi\sim 10$.

The term non-analytic in the quark mass ($m_\pi\sim\sqrt{m_{u,d}}$) proportional to $(\mpi/m_N) \kappa_{0,1}$ in Eq.~\eqref{psEDMres} respects, in principle, the chiral/NDA estimate, but in the HBChPT calculation this term is absent. We can trace this discrepancy back to the use of the pseudoscalar coupling in Eq.~\eqref{pseudoscalar} which breaks chiral symmetry. Repeating the calculation with the relativistic axial-vector pion-nucleon coupling that conserves chiral symmetry, we find that the $(\mpi/m_N) \kappa_{0,1}$ term does not appear. The first non-analytic terms that appear are then proportional to $(\mpi^2/m_N^2) \kappa_{0,1}$, as they should be. We refrain from giving all details, but have confirmed that by using appropriate matching conditions the non-analytic terms appearing in the HBChPT and relativistic calculation (with axial-vector pion-nucleon coupling) are the same\footnote{The chiral symmetry-breaking impact of the pseudoscalar coupling enters through the tree-level, parity-conserving, pion-photon production sub-graphs of the one-loop TVPV diagrams. Matching onto the HBChPT Lagrangian requires introduction of an explicit chiral symmetry-breaking operator. No such operator is required when matching with the relativistic, pseudovector calculation. }. Of course, in the relativistic calculation with pseudovector coupling there still appear analytic terms that are lower order in $m_\pi/m_N$ that may overestimate the dependence of the nucleon EDM on $\gpbz$ and $\gpbo$.

The relativistic results in Eqs. \eqref{psEDMres} and \eqref{FP}
describe short distance physics and can be interpreted as
model-dependent estimates of the size of the LECs $\bar d_0$ and
$\bar d_1$. Due to the large value of the anomalous magnetic
moments, they come to dominate the model estimates of $\bar d_0$ and
$\bar d_1$, given in Ref. \cite{ji}. To see this dominance
explicitly, one may substitute the NDA estimate in
Eq.~(\ref{eq:nda1}) into the expression for $d_0$ given in
Eq.~(\ref{psEDMres}), yielding an estimate for this short distance
contribution that is larger than the NDA estimate in Eq.
\eqref{eq:nda2} by a factor of $\kappa_1$ times a number of order
unity. On general grounds, there is no {\em a priori} reason to
expect the short distance contribution to be enhanced by the
isovector magnetic moment. Moreover, as we have shown above, the
reliance on a one-loop calculation to arrive at such a conclusion
cannot be reproduced in an EFT with consistent power counting and
is, therefore, questionable. On the other hand, the remaining
contributions to the nucleon EDMs calculated in Ref. \cite{ji} not
generated by relativistic, meson theory loops are of similar size as
the NDA estimate in Eq. \eqref{eq:nda2} and, thus, appear quite
reasonable from the EFT standpoint. More generally,  we stress that
the more conservative result of Eq. \eqref{eq:nda2} should be used
to set bounds on scenarios of BSM physics such as left-right
symmetric models.

\section{Implications and Conclusions}
\label{sec:imply}

From the foregoing discussion, it is evident that any phenomenological constraints obtained from experimental limits on $d_n$ and the results in Eq.~(\ref{psEDMres}) may be as much as an order of magnitude too stringent. In HBChPT, the loop contributions proportional to $\kappa_{0,1}$ arise at NNLO in the
$Q/m_N$ expansion, while $\kappa$-independent contributions start off at NLO. For the isoscalar EDM, the NNLO contribution is numerically comparable to the NLO term due to the large magnitude of $\kappa_1$, whereas for the neutron the NLO contribution is absent and the NNLO suppression is more severe. In terms of the \lq\lq left-right symmetric" four-quark interaction (\ref{eq:lreff0}), the NLO loop contributions and the chiral/NDA estimates for the short distance LEC are numerically comparable although the loops can be expected to be somewhat smaller due to to a $\mpi/m_N$ suppression. Consequently, we see no reasonable way around the $m_\pi/m_N$ suppression associated with long-distance contributions containing $\gpbo$ or the chiral/NDA estimate for the short distance contribution, both of which should apply to the analysis of EDM constraints on the LRSM. For other scenarios that lead to a different subset of the $d=6$ CPV operators, such as the Weinberg operator or chiral invariant four-quark operators, the loop contributions associated with $\gpbo$ will be further suppressed compared to the short distance contributions.
On the other hand, a scenario in which the isovector quark chromo-EDM is the dominant $d=6$ CPV source will have a similar phenomenology as the LRSM\footnote{Deviations from these expectations might appear in the EDMs of multi-nucleon systems \cite{Lagdim6}.}.

For the LRSM with spontaneous CPV, the relaxed constraints may allow for consistency between EDMs, the kaon CPV parameter $\epsilon$, and a right-handed $W$-boson ($W_R$) with mass of a few TeV rather than the lower bound of 10 TeV quoted in Ref.~\cite{Zhang:2007da} (see the right panel of Fig. 8 for an illustration of the impact of relaxed EDM constraints). In this case, discovery of the $W_R$ at the Large Hadron Collider would still be possible, though it would likely imply that a second neutral CP-even Higgs boson would be too heavy to be observed.

Similar conclusions apply to the analysis of Ref.~\cite{Ng:2011ui} that considered the implications of $d_n$ limits for the P-conserving, T-violating neutron decay correlation ${\vec J}\cdot{\vec p}_e\times{\vec p}_\nu$ with a coefficient denoted by $D$. Separating out the final state interactions ($D_f$) that can mimic {\em bona fide} T-violation ($D_t$), the authors of Ref.~\cite{Ng:2011ui}  use the work of Ref.~\cite{ji} that, in turn, relied one the relativistic meson theory calculation of Ref.~\cite{He:1992db}, to conclude that $|D_t/\kappa| < 10^{-7}$, where $\kappa$ denotes the combination of Fermi and Gamow-Teller matrix elements, $g_A$, and the vector coupling $g_V$ that enters the correlation.  This bound lies well below the computed final state interaction contribution $|D_f\/\kappa| \sim 10^{-5}$ with an uncertainty at the $10^{-7}$ level. In this case, it would not be possible to disentangle final state contributions from possible LRSM contributions for experiment performed with $10^{-7}$ level sensitivity. However, applying the relaxed EDM constraint we obtain here would lead to $|D_t/\kappa| < 10^{-6}$, opening a window for a future $D$-correlation experiment to observe a non-vanishing LRSM effect.

\acknowledgements
We thank Ulf Mei\ss ner for helpful comments on the manuscript. This work is supported in part by the DFG and the NSFC through funds
provided to the Sino-German CRC 110 ``Symmetries and the Emergence of
Structure in QCD" (JdV), and by
the US Department of Energy under contracts DE-AC02-05CH11231 (EM), DE-FG02-08ER41531 (C-YS, HP and MJRM), and DE-SC0011095 (C-YS and MJRM) .

\end{document}